\documentclass[10pt,journal,letterpaper,compsoc]{IEEEtran}
\usepackage{amsmath}
\usepackage{amssymb}
\usepackage{graphicx}
\usepackage{subfigure}
\usepackage{a4wide}
\usepackage{algorithm}
\usepackage{algorithmic}
\usepackage{amsfonts}
\usepackage{color}
\usepackage{framed}
\usepackage{verbatim}

\makeatletter
\def\ps@headings{%
\def\@oddhead{\mbox{}\scriptsize\rightmark \hfil \thepage}%
\def\@evenhead{\scriptsize\thepage \hfil \leftmark\mbox{}}%
\def\@oddfoot{}%
\def\@evenfoot{}}
\makeatother
\usepackage{todonotes}
\definecolor{lemonchiffon}{rgb}{1.0, 0.98, 0.8}

\begin{document}
\title{Cascading Failures in Smart Grids under Random, Targeted and Adaptive Attacks}

\author{\IEEEauthorblockN{
Sushmita Ruj$^\dag$ (\emph{Senior Member, IEEE}) and Arindam Pal$^\ddag$ (\emph{Senior Member, IEEE})\\}
\IEEEauthorblockA{
$^\dag$ UNSW Sydney, New South Wales, Australia. Email: sushmita.ruj@gmail.com \\
$^\ddag$ Data61, CSIRO, Sydney, New South Wales, Australia. Email: arindamp@gmail.com
}}

\maketitle{}

\begin{abstract}
A smart grid consists of two networks: the power network and the communication network, which are interconnected by edges spanning across the networks. We model smart grids as complex interdependent networks, and study targeted and adaptive attacks on smart grids for the first time. Due to attack on one network, nodes in the other network might get isolated, which in turn will disconnect nodes in the first network. Such cascading failures can result in disintegration of either or both of the networks. Earlier works considered only random failures. In real life, an attacker is more likely to compromise nodes selectively.

We study cascading failures in smart grids, where an attacker selectively compromises the nodes with probabilities proportional to their degrees, betweenness, or clustering coefficient. This implies that nodes with high degrees, betweenness, or clustering coefficients are attacked with higher probability. We mathematically and experimentally analyze the sizes of the giant components of the networks under different types of targeted attacks, and compare the results with the corresponding sizes under random attacks. We show that networks disintegrate faster for targeted attacks compared to random attacks. A targeted attack on a small fraction of high degree nodes disintegrates one or both of the networks, whereas both the networks contain giant components for random attack on the same fraction of nodes. An important observation is that an attacker has an advantage if it compromises nodes based on their betweenness, rather than based on degree or clustering coefficient.

We next study adaptive attacks, where an attacker compromises nodes in rounds. Here, some nodes are compromised in each round based on their degree, betweenness or clustering coefficients, instead of compromising all nodes together. In this case, the degree, betweenness, or clustering coefficient is calculated before the start of each round, instead of at the beginning. We show experimentally that an adversary has an advantage in this adaptive approach, compared to compromising the same number of nodes all at once.  
\end{abstract}

\textbf{Keywords}: Complex Networks, Percolation Theory, Smart Grids, Cascading Failures, Degree, Betweenness, Clustering Coefficients, Random, Targeted and Adaptive Attacks.

\section{Introduction}
\label{sec:intro}
Power grids have suffered severe failures in the past. 
The black out of Northern US/Canada and that of Italy in 2003 affected the lives of millions of people and resulted 
in huge monetary losses. 
More recently, the largest blackout in the world occurred in India in July 2012.
The complete shutdown of the  Northern, Eastern, and Northeastern power grids in the country affected over 620 million people.
Such calamities could have been avoided, if the power grid functioned properly. 
In order to ensure that the electric grid functions smoothly, it is important that the control
information is collected and transmitted in an orderly fashion, and the existing systems be highly automated. 
Smart grids are next generation electricity grids, in which the power network and the communication network work in tandem. Smart grids promise to fulfill this vision by synchronizing the power network with the communication network. The idea is to replace the existing \emph{SCADA (Supervisory Control and Data Acquisition)} system  by an intelligent and automatic communication network. 

The power network consists of power plants, generation and distribution stations, whereas the communication network
consists of sensors attached to appliances to collect information, aggregator sensors to aggregate information and smart meters for monitoring and billing. 
The smart meters in home area networks, building area networks, and
neighborhood area networks are responsible for aggregating, processing and transmitting data and control information for proper functioning of the smart grid. 
The question is how to make such a network robust and fault-tolerant. 
Researchers have addressed smart grid architectures \cite{B10} and the problem of cascading failures \cite{CPS11}, 
in which a small fault propagates throughout the network and affects a large part of the network. 
Most of the current techniques and models use concepts from distributed systems. 
However, because of the large size of smart grids and their unique properties, new models, 
interconnection patterns, and analysis techniques are required to increase  the robustness of networks.

Recently, Huang \emph{et al.} \cite{HWRSN13} initiated the study of modeling and analyzing smart grids using
interdependent  complex networks.   
A smart grid can be thought of as two complex networks, which are interconnected. 
The question is how to make this network robust and fault tolerant. 
In order to provide a solution, we have to understand what kind of faults and attacks can take place and 
how faults propagate in the network.  
The failure of nodes in one network results in the disruption of the other network, which in turn affects the first network. 
This type of failure propagates in a cascading manner and was the main reason for the blackouts in the US and in India. 
To understand this \emph{cascading failure}, we need to study the structure of the networks.
In this paper, we model and study smart grids as complex networks and show the effect of cascading failure, when adversaries compromise nodes in the network. 

Though cyber-security issues have been studied in details \cite{WL13}, modeling the network
in order to make it resilient still needs lot of research.  
The main contribution of this paper is to study the effect of targeted and adaptive attacks in smart grids, 
in which the attacker selectively disrupts communication nodes. 
{\textcolor{black}{This is one of the early works on targeted and adaptive attacks on smart grids using complex network model.}}
We argue that an adversary is more likely to attack selected high degree (or betweenness
or clustering coefficient) nodes, rather than attacking nodes randomly. 
As an example, we consider the recent Stuxnet worm \cite{STUX13} which was targeted on Siemens PCs and caused large-scale destruction to industrial control systems.  
Yagan \emph{et al.} \cite{YQZC12} studied cascading failures in cyber-physical systems. 
They studied different interdependent Erdos-Renyi (ER) networks \cite{N10}, but they did not consider scale-free networks, which are used to 
represent power and communication networks. 
Till date, all works \cite{HWRSN13,HWSN13,YQZC12,RP14} on complex networks models of smart grid have considered only random attacks.
In the preliminary version \cite{RP14} of this paper, we addressed targeted attacks. We analyzed the sizes of giant components in each network, under targeted attacks. However, nodes were compromised based only on the degree of the node, i.e. high degree nodes were compromised with high probability. Betweenness and clustering coefficients were not considered. Adaptive attacks were also not considered in the preliminary version. 
Huang \emph{et al.} \cite{HWSN13} addressed the cost of maintaining such networks by analyzing the number of support links between networks.
Whereas increasing the support links might make the interdependent networks stronger, large number of support links
imply higher cost of maintenance. They suggested that smart grids should have some nodes which are connected to power nodes (also called \emph{operation centers})
and the rest of the nodes are \emph{relaying nodes}. Using such a model, they studied the resilience of the network under random attacks.
According to their model, each control node is linked to $n$ power nodes and each power node is operated by $k$ operation centers.

{\textcolor{black}{
Interdependent network have been studied in the context of cyber-physical systems in general by \cite{HWSN15}. They studied cascading failures in interdependent networks.
The papers studies random attacks on the networks. 
They calculate percolation thresholds for interdependent networks using extensive experiments. 
Interdependent networks have also been addressed for complex contagion in \cite{YG13}. 
}}

\subsection{Problem statement and our contribution}
\label{sec:contribution}
We model the smart grid as a complex interdependent network consisting of two networks, the power network and the communication network. 
Both the power network and the communication network are scale-free (SF) networks, where the degree distribution follows the power law, 
$p_k \propto k^{-\alpha}$, where $p_k$ is the fraction of nodes of degree $k$ and $\alpha$ is the power-law parameter specific to the network. 
Scale free networks are a type of random graphs which commonly arise in many practical cases, like social, biological networks, the Internet, power grids, to name only a few. Another type of network, which is often studied is the Erdos-Renyi (ER) network, denoted by $G(n,p)$, where $n$  is the number of nodes and $p$ is the probability that an edge exists between two nodes. 
Support links are randomly assigned from one network to another, such that a power node is controlled by multiple communication nodes, and
functions properly as long as at least one such link exists. 
In our model, we consider targeted attacks on the communication network. 
We mathematically analyze the effect of cascading failure for this type of attack and 
find out the sizes of giant components when nodes are compromised. 

{\textcolor{black}{
We compare the following attack models: random attacks, targeted attacks, and a combination of targeted and random attacks.
We show that 
an adversary has a definite advantage if it compromises nodes selectively. 
A simple example is that if an adversary wants to launch a terrorist attack, it would like to plant as few bombs as possible, while
maximizing the damages. Thus, the adversary has to compromise nodes selectively. Critical node detection is an interesting
problem, which has been studied in literature in many contexts. 
As pointed out in \cite{NST13,SMBDC14}, detecting critical nodes in an interdependent power grid is a  NP-complete problem. 
}

In this paper we have considered the following strategies for compromising nodes selectively:
the attacker might consider either the degree or betweenness or clustering coefficient and compromise nodes with high degree (or betweenness or clustering coefficient). \emph{Betweenness} of a node is the number of shortest paths that pass through a node. \emph{Clustering coefficient} is the number of common neighbors of two nodes which are neighbors themselves. A formal definition appears in 
Section \ref{sec:background}.
In targeted attack, the adversary compromises a node with a probability proportional to the degree (or betweenness or clustering coefficient) of the node. 
We show that, from the point of view of the adversary, compromising nodes with probability proportional to the betweenness is better than compromising
nodes either randomly or by compromising nodes with probability proportional to the degree or clustering coefficients. 
We compare our results for a combination of SF-SF networks and ER-ER networks. We show that SF-SF networks are more vulnerable to targeted attacks than ER-ER networks. We also analyze the average path length under targeted attacks. 

Next we compromise nodes adaptively. This means that instead of selecting all nodes to be compromised at the start, 
we compromise nodes (based on degree/betweenness/clustering coefficients) in rounds. This implies that an attacker compromises a few nodes in round one and then depending 
upon the cascading effect on the two networks compromises another set of nodes in the second round and so on. In this case, we show that an adversary has an advantage for adaptive attacks over non-adaptive attacks. Adaptive attacks based on betweenness result in smaller giant component compared to Adaptive attacks based on degree.

Our main conclusion is that by launching a targeted attack, an adversary can disrupt significant part of the network. 
For a large network, compromising about 2.2\% of the network can disrupt either of the networks under targeted attack, whereas under random attack, the networks are still connected and work smoothly. 

{\textcolor{black}{
We observe that after targeted attacks, the size of the giant component in Erdos-Renyi networks can be twice as large as that in Scale-Free  networks. 
} 

\subsection{Organization}
The paper is organized as follows. 
Related works are presented in Section \ref{sec:related}. Preliminary material on complex networks is given in Section \ref{sec:background}. Network model and attack model are presented in Section \ref{sec:model}. The basic technique for computing the size of the giant component is described in Section \ref{sec-giantcomponent}. Cascading failure is mathematically analyzed in Section \ref{sec:analysis}. 
In Section \ref{sec:experimental}, we present experimental results to understand our model and make some conclusions. 
We conclude in Section \ref{sec:conclusion} with directions for future work.

\section{Related works}
\label{sec:related}
Smart grid communication and network architecture have been widely studied in \cite{B10,WXK11,LXLLC12}.
Most smart grid literature concentrate on distribution of power \cite{YLR12}, balancing supply and demand \cite{RWJSL10}, 
detecting and predicting faults \cite{CPS11}, designing network architecture which are fault tolerant \cite{YQZC12}. 
The bulk of literature on fault tolerance address cyber-physical systems in general \cite{YQZC12} and use general models and techniques 
of distributed systems.

{\textcolor{black}{
The cyber security requirements of smart grids have been outlined by the National Institute of Standards and Technology (NIST) \cite{NIST10}.
There has been extensive research in the security and privacy of smart grids  in recent years. Surveys can be found in \cite{LXLLC12,WL13}. 
The main problems that have been explored include secure and privacy preserving data aggregation in smart meters which have been studied in (\cite{LLL10,LLLLS12,RN13,RD10}). Privacy preserving smart metering has been addressed in \cite{AC11,MDFSI12}. Access control  of smart grid data have been presented in \cite{RN13,BKAA09}. 
Data authentication has also been studied by Fouda \emph{et al.} \cite{FFKLS11}, Kgwadi-Kunz \cite{KK10} and Lu \emph{et al.} 
\cite{LLSS13, LLLLS12}. 
}}

{\textcolor{black}{
A coordinated multi-switch attack was proposed in \cite{LCZKB14}. The opponent is able to control many switches in the power system. Using dynamical systems, the authors show how to launch an attack by using multiple beakers. Other switching attacks have been studied in \cite{LMKZB13,LKZB12}. 
}}

{\textcolor{black}{
Fault tolerance in power grids has been studied widely in the past. 
Efficient methods of load distribution to prevent cascading failures have been studied in \cite{CDW10}. 
Load Redistribution (LR) attack was developed and studied in \cite{YZKR11} by analyzing the extent of damage to power grid operation. 
 The  power grid has been modeled as a graph and the robustness has been stuided in \cite{SN10}. 
By extensive mathematical analysis, it has been  estimated that the disturbance levels the system can accept before a few overload nodes resulting a large blackout. 
Static overload failure was discussed in \cite{CPS11}. Optimization techniques are used and a distance-to-failure algorithm was proposed to predict the weak points in power grid. 
They proposed an attack model describing the main goal of LR attack, and then based on that indicated the theory and criterion of protecting the system from LR attack.
To decrease the impact of overload cascading, \cite{RKM11} proposed a model to focus on the analysis of tripping of already overloaded lines. By simulation on a real-world power grid structure, it is shown that controlled tripping of overloaded lines leads to significant mitigation of cascading failure.
}

The study of the model, analysis of the network structure, increasing the robustness
of power grids have been studied using complex networks. 
Here, electric distribution stations, 
transmission stations, generation centers are modeled as nodes. Two nodes are connected by a link, if there is power flow from
one node to the other. The structure of the underlying graph has been widely studied to find the effect of node failures. 
When certain nodes fail (or are attacked), the links incident on these nodes are disrupted. This affects other nodes, whose links fail in turn. 
Such failures propagate in a cascading manner throughout the network. Thus, a small fraction of nodes can disrupt a large part of the network.
It has been shown that the graph structure underlying a power grid follows a power law distribution \cite{N10}. 
An extensive survey appears in \cite{PA13}. 
 
Although, complex networks have been widely used to study different networks like social networks, biological networks, citation networks, 
power networks \cite{N10}, smart grid networks have not been widely studied. 
Huang \emph{et al.} \cite{HWRSN13} introduced the study of smart grids using complex interdependent networks, in which the power network and the communication network are modeled as individual networks which have scale-free property. 
The links connecting nodes within a network are called \emph{intralinks}. 
The networks are connected to each other via links (also called \emph{interlinks}), 
such that a power node depends on communication nodes and vice versa. 
Such a network is called an \emph{interdependent network}. 

Interdependent networks were introduced by Buldyrev \emph{et al.} \cite{BPPSH10}. They studied the effect of failure cascades in such networks.
The failure of a few nodes in the communication network will affect nodes in the power network, which will further affect nodes in the communication network.
Thus, failures propagate in cascades till a steady state is reached or when either or both of the  networks disintegrate. 
We say that a network \emph{disintegrates} if there are no \emph{giant components} in the network. A giant component is a connected component of size $\Theta(N)$, where $N$ is the number of nodes in the network. Since then, a number of researchers have analyzed interdependent networks. 

The initial study by Buldyrev \emph{et al.} \cite{BPPSH10} studied the case where the two networks are of the same size, and there is a one-to-one correspondence between nodes which are joined by an interlink. 
Shao \emph{et al.} \cite{SBHS11} studied multiple support interlinks, where a node in the power network was connected to multiple nodes in the communication network and vice-versa. 
Most of the results have been analyzed experimentally, because closed-form analytical solutions are difficult to obtain. 
A special case of support links, where nodes having identical degree are connected across networks was studied  in \cite{BSC11}. 
It has been observed in all these cases that interdependent systems make the network much more vulnerable to attacks, compared to a single network. 

A well-known result in complex networks is that, randomly removing 95\% of the nodes in the Internet (which is a scale-free network) can still result in a connected network. However, strategically removing even 2.5\% of the nodes can disrupt the whole network \cite{D07}. 
Such a result motivates us to study the effect of targeted attacks on smart grids. 
In case of smart grids, an adversary is more likely to compromise nodes of strategic importance like hubs, than nodes of low degree. 
Thus, selective attacks give substantially different results compared to random attacks. 

Targeted attacks on interdependent networks has been studied in \cite{HGBHS11}. {\textcolor{black}{The attacker chooses the nodes with probability proportional to the degree. It follows that a high degree node has a higher probability of being attacked.}} 
{\textcolor{black}{
Our work is significantly different from theirs in the following respects. 
(i) In \cite{HGBHS11}, the authors  considered only targeted attacks based on degree. We have compared targeted attacks based on degree, betweenness centrality and clustering coefficients, and also random attacks. 
(ii) The paper \cite{HGBHS11} assumed that the two networks are of the same size and same type (both ER-ER or both SF-SF). So, there is one-one correspondence between the nodes in either parts of the network. We have considered general networks even with unequal number of nodes in the two parts. 
(iii) The main aim of \cite{HGBHS11} was to find the percolation threshold. In this paper, we compare different parameters like size of the giant component, average path length and other parameters under different attack strategies, when a given number of nodes are compromised. 
(iv)	Adaptive attack has been studied in our paper, but not in \cite{HGBHS11}. }}

Instead of studying interdependent networks consisting of two networks, Dong \emph{et al.} \cite{DGDTSH13} studied targeted attacks on a network of networks. 
Zheng and Liu \cite{ZL12} proposed a solution for making a network robust against targeted attacks by suggesting a onion-like structure. 
Here high-degree nodes are present towards the center in clusters and low-degree nodes are present in concentric rings depending upon their degree. 
They analyzed results from power networks. Their technique is however restricted to single networks. 

Ruj and Pal \cite{Ruj2019} discussed different network models of smart grids and their impact on the reliability and availability. They analyzed various techniques to increase the resilience of networks. Zhu et al. \cite{zhu2020methodology} proposed an analytical method, based on complex networks, to assess the risk of the Smart Grid failure due to communication network malfunction, associated with latency and ICT network reliability. The proposed approach is tested on a laboratory-scale communication network. Jiang et al. \cite{jiang2019evolutionary} developed an evolutionary computation based vulnerability analysis framework, which employs particle swarm optimization to search the critical attack sequence. Zuniga et al. \cite{zuniga2020classical} introduced the application of Failure Modes and Effects Analysis (FMEA) method in future smart grid systems in order to establish the impact of different failure modes on their performance. They proposed a reliability based approach that makes use of failure modes of power and cyber network main components to evaluate risk analysis in smart electrical distribution systems. Gupta et al. \cite{gupta2014probabilistic} proposed a probabilistic framework of smart grid power network with statistical decision theory to evaluate system performance in steady state, as well as under dynamical case, and identify the probable critical links which can cause cascade failure. They developed a graphical model using minimum spanning trees to analyze topology and structural connectivity of IEEE 30 bus system.


\section{Background on Networks}
\label{sec:background}
In this section we will define a few terms related to networks, that will be used later. 
The \emph{degree} of a node is defined as the number of edges that are incident on that node. 
\emph{Degree distribution} is a random variable $X$, such that $P(X=d)$ is the fraction of nodes which have degree $d$. 
\emph{Centrality} of a node measures its relative importance in the network. 
It is measured by parameters such as degree, betweenness and clustering coefficient. 
\emph{Betweenness} of a node is the number of shortest paths that pass through the node. 
\emph{Clustering coefficient} is measured by two parameters, local clustering coefficient and global clustering coefficient.
Local clustering coefficients also called transitivity is measures the probability that two neighbors of a vertex are connected. 
More precisely, this is the ratio of the triangles and connected triples in the graph.

{\textcolor{black}{Calculating the betweenness centralities of all the vertices in a graph requires finding the shortest paths between all pairs of vertices on a graph, which takes $\Theta(n^3)$ time with the Floyd–-Warshall algorithm, by modifying it to find all shortest paths between two nodes. On a sparse graph, Johnson's algorithm takes $O(n^2 \log n + nm)$ time. On unweighted graphs, calculating betweenness centrality takes $O(nm)$ time using Brandes' algorithm \cite{brandes2001faster}. Here, $n = |V|$ and $m = |E|$ are the number of vertices and edges of the graph $G = (V, E)$ respectively.}} 
{\textcolor{black}{Existing randomized and parallel algorithms \cite{BP07,BM06,KPST11, MB14}, for calculating centrality measures for large graphs can be used. 
Throughout the paper we assume for simplicity that the adversary has complete knowledge of the network. If complete knowledge is not available, then the adversary might use incremental algorithms \cite{K13} to calculate centrality. In such cases the centrality measures are calculated iteratively, as and when the adversary gradually gains more knowledge of the network.  }} 

A \emph{giant component} in a graph on $n$ vertices is a maximal connected component with at least $cn$ vertices, for some constant $c$. If $c = 0.5$, this means that the giant component should have at least half of the vertices in the graph.
{\textcolor{black}{A giant component gives a measure of the connectivity of a network. For a power grid, if there exists a giant component containing $80\%$ of the vertices, then these vertices can communicate among each other. If all components are small, the connectivity is very limited.}}
{\textcolor{black}{The size of the giant component and its existance becomes important when nodes in a power/communication network are compromised.}

\section{Smart grid model}
\label{sec:model}
We will first discuss the network model and then the attack model. 
\subsection{Network Model}
We consider two interdependent scale free networks, a communication network $N_A = (V_A,\alpha_A)$ and a power network $N_B = (V_B, \alpha_B)$, 
where $n_A = |V_A|$ and $n_B = |V_B|$ are the number of nodes in the communication and power networks, respectively, and $\alpha_A$ and $\alpha_B$ are the power-law coefficients. 
This implies that, $N_A$ has the power law distribution $P_A(k) \propto k^{-\alpha_A}$, which means that the fraction  of nodes
with degree $k$ is $P_A(k)$. 
Similarly, $N_B$ has the power law distribution $P_B(k) \propto k^{-\alpha_B}$. 
We assume that there are more communication nodes than power stations, which implies that $n_A >n_B$. 

The interlinks, also called  support links \cite{SBHS11} are directed edges from one network to the other. 
We assume that a communication link supports one power station and is powered by one power node, meaning that both the in-degree and out-degree of a
communication node is one. 
A power node is controlled by multiple communication node and supplies power to multiple communication nodes, 
meaning both the in-degree and out-degree of a power node is at least one. 
Links are assigned randomly from the communication network $N_A$ to the power network $N_B$. 

{\textcolor{black}{
We have considered a simple model for understanding the dynamics of the networks. In future, we will extend our model and include other relevant parameters. The following papers \cite{HWRSN13,HWSN13,BPPSH10,HWSN15} have previously considered this simple model. }}
Let $\tilde{k}_A$ denote the support degree of a node in Network $A$. 
This implies that there are $\tilde{k}_A$ nodes in $N_B$, that support a node in $N_A$. 
Let $\tilde{P}_A(\tilde{k}_A)$ denote the degree distribution of support links from $N_B$ to $N_A$.
$\tilde{P}_B(\tilde{k}_B)$ can be defined analogously. 
From the structure of the network, $\tilde{k}_A$ is equal to one for all nodes in $N_A$. 

To calculate the degree distribution $\tilde{P}_B(\tilde{k}_B)$, we note that the problem of assigning 
support links from $N_A$ to $N_B$ is equivalent to assigning $n_A$ balls randomly into $n_B$ bins. 
If $X_i$ denotes the random variable that counts the number of balls in bin $i$, 
then,  
\begin{center}
$Pr[X_i = k] = \binom{n_A}{k} \left(\frac{1}{n_B}\right)^k \left(1-\frac{1}{n_B}\right)^{n_B - k}$.
\end{center}
Thus, the degree distribution $\tilde{P}_B(\tilde{k}_B)$ follows Binomial distribution with parameters $Bin(n_A,\frac{1}{n_B})$. 

\subsection{Attack Model}
{\textcolor{black}{
We consider targeted attack on communication network.
The attacker chooses the nodes with probability
proportional to the degree or betweenness or clustering coefficient. It follows that a high
degree/betweenness/clustering coefficient node has a higher probability of being attacked.
}}
Targeted attacks are more likely to arise in real-world situations, as we have seen during the recent Stuxnet attack. 
Attacking the high degree node is also intuitive, since disrupting the high degree nodes result in more connections being disrupted, 
thus disrupting the network.

We also consider adaptive attacks which we have discussed in Section \ref{sec:adaptive_attack}.

\begin{figure}[ht]
\centering
\includegraphics[scale=1.0]{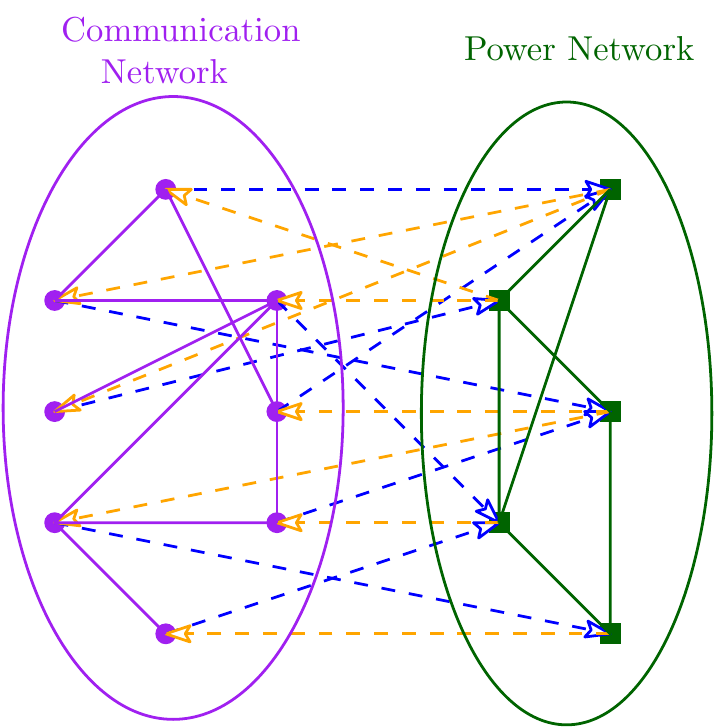}
\caption{The smart grid as an interdependent complex network.}
\label{CommPower}
\end{figure}
 
A vertex can be deleted from the graph in any of these cases.

\begin{enumerate}
	\item If the vertex is attacked.
	\item If the vertex becomes isolated.
	\item If the vertex is not attacked, but all its support nodes on the other network has been attacked.
\end{enumerate}

Note that due to this kind of cascading failure of nodes, many more nodes will be compromised. This is different from the normal scenario, where only the attacked nodes are compromised.

\subsection{Notations}
We have used the notations in Table \ref{table:notations} throughout the paper. 
Figure \ref{CommPower} shows a smart grid as an interdependent network. 

\begin{table}[t]
\label{table:notations}
\centering
\normalsize
\caption{Notations}
\label{notations1}
\scalebox{0.9}{
    \begin{tabular}{|c|p{7cm}|}
        \hline
	$N_A$ & Communication network\\
        \hline
	$N_B$ & Power network\\
        \hline
	$n_A, n_B$ & Number of nodes in $N_A$ and $N_B$ \\
        \hline
	$k$ & Degree of a node\\
	\hline
	$P_A(k)$ & Degree distribution of  communication network \\
	\hline
	$P_B(k)$ & Degree distribution of power network \\
	\hline
	$\tilde{P}_A(\tilde{k}_A)$ & Degree distribution of support degree of a node in $N_A$\\
	\hline
	$\tilde{P}_B(\tilde{k}_B)$ & Degree distribution of support degree of a node in $N_B$\\
	\hline
	$\mathcal{G}A_n$ & Giant component of $N_A$ at stage $n$\\
	\hline
	$\mathcal{G}B_n$ & Giant component of $N_B$ at stage $n$\\
	\hline
	$q_{A_k}$ & Probability of a node having excess degree $k$ (\emph{i.e.,} total degree $k+1$) in $N_A$ \\
	\hline
	$q_{B_k}$ & Probability of a node having excess degree $k$ (\emph{i.e.,} total degree $k+1$) in $N_B$ \\
	\hline
	$r_{A_n}$ & Fraction of removed nodes in $N_A$ at stage $n$, due to removal of nodes in $N_B$ at stage $n-1$ \\
	\hline
	$r_{B_n}$ & Fraction of removed nodes in $N_B$ at stage $n$, due to removal of nodes in $N_A$ at stage $n-1$ \\
	\hline
	$\mu_{A_n}$ & Fraction of functional nodes in $N_A$ at stage $n$\\
	\hline
	$\mu_{B_n}$ & Fraction of functional nodes in $N_B$ at stage $n$\\
       \hline

    \end{tabular}
    }
\end{table}

\section{Calculating giant component upon random removal of vertices}
\label{sec-giantcomponent}
In this section we will study the effect of random node compromise on a single network. We will see how this result can be used for analyzing attack on interdependent networks. 
Let us consider a network $N$ having a degree distribution $P(k)$. 
Let $\phi$ be the fraction of nodes left after random removal of nodes. 
Let $u$ be the probability that a vertex is not connected to the giant component via a particular neighbor. 
If the  vertex has degree $k$, then average probability that it is not in the giant component is 
\begin{equation}
g_0(u) = \sum_k P(k) u^k, 
\end{equation}
where $g_0(z) = \sum_k P(k) z^k$, is the generating function for the degree distribution. 
Hence, the probability that a vertex belongs to a giant component is $1-g_0(u)$. 
However, the vertex itself is present with a probability $\phi$. 
Thus fraction of nodes in the giant component is 
\begin{equation}
\mu_N = \phi(1-g_0(u))
\end{equation}

In order to calculate the value of $u$, we note that a node $i$ is not in the giant component if it is either removed, 
or it is present but not connected to the giant component via any of its neighbors. 
The first condition happens with probability $1-\phi$ whereas the second condition happens with probability $\phi u^k$. 
Since node $i$ can be reached following an edge, 
the value of $k$ follows the excess degree distribution 
\begin{equation}
q_k = \frac{(k+1)q_{k+1}}{\langle k \rangle},
\end{equation}
where $\langle k \rangle$ is the average degree of the network. 
Thus, averaging over this distribution we get
\begin{equation}
\begin{split}
u & = \displaystyle \sum_{k=0}^{\infty}q_k(1-\phi + \phi u^k) \\
& = 1 - \phi + \displaystyle \sum_{k=0}^{\infty} q_k u^k\\
& = 1 - \phi + \phi g_1(u),
\end{split}
\end{equation}
where 
\begin{equation}
g_1(z) = \displaystyle \sum_{k=0}^{\infty} q_kz^k
\end{equation}
is the generating function for excess degree distribution. 
A detailed analysis can be found in \cite{N10}.

\section{Modeling cascading failure due to targeted attack on communication network}
\label{sec:analysis}
We first analyze the targeted attack on communication network and then show how the failure propagates across the interdependent networks in stages. This is represented in Figure \ref{fig:cascade}.

\begin{figure*}
\begin{centering}
\includegraphics[width=6.0in]{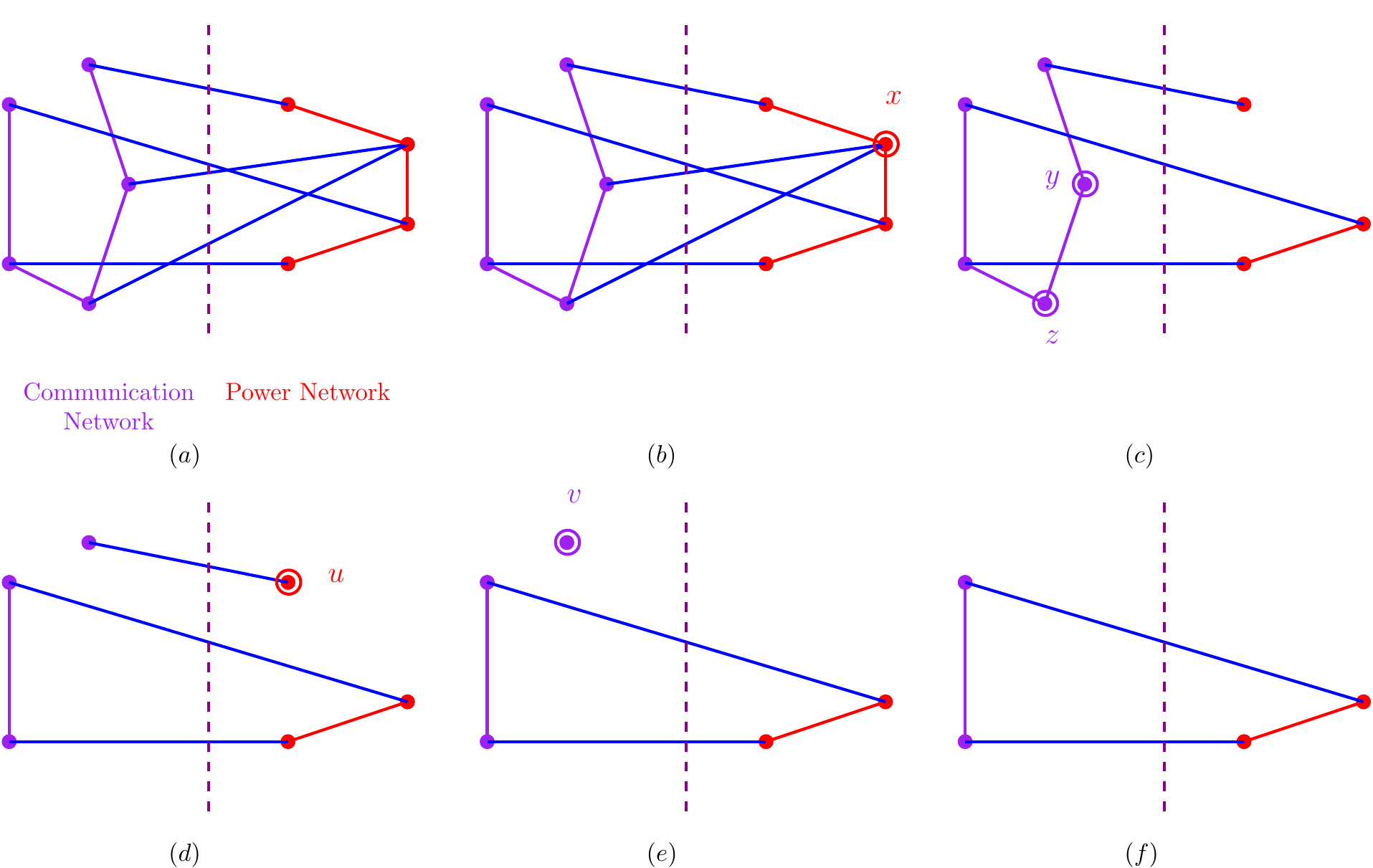}
\caption{Cascading failures in interdependent smart grid networks. The faulty nodes are shown as double circles.}
\label{fig:cascade}
\end{centering}
\end{figure*}

\subsection{Stage I: Targeted attack on the communication network}
\label{subsec:target-attack-communication-node}
We consider three types of targeted attacks. 
A node is removed with probability proportional to its degree or betweenness or clustering coefficient. 
We assume that the attacker uses only one of these centrality measures for node compromise, but not a combination of all three. 
Let $\phi_k$ be the probability that a node $i$ of centrality measure $k$ (either degree or betweenness or clustering coefficient of 
node $i$) is not removed. Clearly,
\begin{align*}
\phi_{k} &= 1 - \frac{centrality(i)}{\sum_{v \in V_A} centrality(v)} \\
\end{align*}
For example, $deg(i) = A k^{-\alpha_A}$ is the degree of node $i$ and $m_A$ is the number of edges in $N_A$.
If the attacker decides to remove nodes with high degree then, 
\begin{align*}
\phi_{k} = 1-\frac{A k^{-\alpha_A}}{2m_A}.
\end{align*}
We note that $\alpha_A = 0$ represents random removal of nodes. 

We will first calculate the size of the giant component $\mathcal{G}A_1$. 
Let $u$ denote the average probability that a node is not connected to the giant cluster via one of its neighbors. 
Consider a node of degree $k$. 
Probability that it is not connected to the giant component via any of its neighbors is $u^k$.
\begin{center}
Probability of it being in the giant component = \\
Probability that it is not attacked $\cdot$ probability that one of its neighbors is in the giant component.
\end{center}
Thus, probability of it being in the giant component is $\phi_k(1-u^k)$. 
Averaging over the degree distribution $P_A(k)$, we can calculate the fraction of nodes in the initial giant component as
\begin{equation}
\begin{split}
\mu_{A_1} &= \sum_{k=0}^{\infty}P_A(k)\phi_k(1-u^k) \\
&= \sum_{k=0}^{\infty}P_A(k)\phi_k - \sum_{k=0}^{\infty}P_A(k)\phi_k u^k \\
& = f_0(1) - f_0(u), 
\end{split}
\end{equation}
where, 
\begin{equation}
f_0(z) = \sum_{k=0}^{\infty}P_A(k)\phi_k z^k.
\end{equation}

We will now show how to calculate $u$. A node is not connected to the giant component when either of the following cases arise. 
\begin{itemize}
\item The node is attacked and thus removed, 
\item The node is present, but not connected to any node in the giant component.
\end{itemize}

Let $k$ be the excess degree of a neighboring node. The original degree of a node is one more than the excess degree, \emph{i.e.,} $k+1$ \cite{N10}. 
Probability that a neighbor is removed is $1 - \phi_{k+1}$. 
Probability that a neighbor is present, but the node itself is not present in the giant component is $\phi_{k+1}u^k$. 

Hence using \cite{N10}, $u$ can be calculated as, 
\begin{equation}
\begin{split}
u & =  \displaystyle \sum_{k=0}^{\infty} q_{A_k}(1 - \phi_{k+1} + \phi_{k+1}u^k)\\
& = 1 - f_1(1) + f_1(u), 
\end{split}
\end{equation}
where, 
\begin{equation}
f_1(z) =  \displaystyle \sum_{k=0}^{\infty} q_{A_k} \phi_{k+1}z^k. 
\end{equation}
Note that $q_{A_k}$, the probability of a node having excess degree $k$ in $N_A$ can be given by $q_{A_k} = \frac{(k+1)P_A(k+1)}{\langle k_A \rangle}$ \cite{N10}. It can be seen that, $\sum_{k=0}^{\infty} q_{A_k} = 1$. 
Substituting the value of $q_{A_{k+1}}$, the value of $f_1(z)$ can be calculated as, 
\begin{equation}
\begin{split}
f_1(z) & = \displaystyle \sum_{k=0}^{\infty} \frac{(k+1)P_A(k+1)}{\langle k_A \rangle} \phi_{k+1} z^k\\
& = \frac{1}{\langle k_A \rangle}  \displaystyle \sum_{k=1}^{\infty}k P_A(k) \phi_k z^{k-1}, 
\end{split}
\end{equation}
where, $\langle k_A \rangle$ is the average degree of nodes in $N_A$. 
We observe that, 
\begin{equation}
f_1(z) = \frac{f_0'(z)}{g_{A_0}'(1)}, 
\end{equation}
where $g_{A_0}(z)$ is the generating function, 
\begin{equation}
g_{A_0}(z) = \displaystyle \sum_{k=0}^{\infty} P_A(k)z^k.
\end{equation}

\subsection{Stage II: Effect of cascading failure on the power network}
\label{subsec:stageII-powernetwork}
{\textcolor{black}{In a power grid, the effect of communication network is not so pronounced. However, in a smart grid the communication and power networks reinforce each other.
}}
{\textcolor{black}{Due to attack on nodes in the communication network, some nodes in the power network might be affected. This happens for smart grids, which are interconnected networks.  }}
A node in the power network $N_B$ is functional, if 
a node in $N_B$ has at least one support link from $N_A$.
Initially, at stage II all nodes in $N_B$ are in the giant component. 
We consider all those nodes which are supported by nodes not in $\mathcal{GA}_1$. 
Such nodes will not remain functional because they will be cut off from the communication network. 
Probability that a node is not in the giant component $\mathcal{G}A_1$ is $1 - \mu_{A_1}$. 
Suppose, a node is supported by $\tilde{k}_B$ nodes in $N_A$. 
Probability that the $\tilde{k}_B$ neighboring nodes are not in  $\mathcal{G}A_1$ is  $(1 - \mu_{A_1})^{\tilde{k}_B}$. 
Fraction of nodes in $N_B$ disconnected due to attack on $N_A$ is given by,  
\begin{equation}
r_{B_2} = \displaystyle \sum_{\tilde{k}_B=0}^{\infty} \tilde{P}_B(\tilde{k}_B) (1 - \mu_{A_1})^{\tilde{k}_B}
\end{equation}

The fraction of nodes remaining in $N_B$ is given by $1 - r_{B_2}$. 
This is similar to the random removal of vertices. 
The fraction of nodes in the resulting giant component can be calculated by the technique in 
as 
\begin{equation}
\mu_{B_2} = (1 -r_{B_2}) (1 - g_{B_0}(u)),
\end{equation}
where,
\begin{equation} 
 u = 1 - \phi + \phi g_{B_1}(u),
\end{equation}
\begin{equation}
g_{B_0}(u) = \displaystyle \sum_{k=0}^{\infty} P_B(k) u^k 
\end{equation} and
\begin{equation}
g_{B_1}(z) = \displaystyle \sum_{k=0}^{\infty} q_{B_k}z^k.
\end{equation}

\subsection{Stage III: Cascading failure in the communication network}
\label{subsec:stageIII-commnetwork}
We will now study the effect of cascading failure in the communication network, due to the failure in power networks. 
Each node in $N_A$ is supported by only one link from the power network. 
If a node in $N_B$ fails, then the communication node it supports, also fails. 
{\textcolor{black}{
We have assumed a simple interconnection pattern for ease of analysis. For more complex interconnection patterns, only the links connecting the failed node in $N_B$ are disrupted. }}
The fraction of nodes in $N_A$ which fail due to failure of node in $N_B$ is given by, 
\begin{equation}
r_{A_3} = \sum_{\tilde{k}_A=0}^{\infty} \tilde{P}_A(\tilde{k}_A) (1-\mu_{B_2}). 
\end{equation}

We can consider that these nodes are randomly removed in $N_A$ and find the giant component resulting due to this removal of nodes. 
The fraction of nodes in the giant component which result from this random compromise is calculated as shown in Section \ref{sec-giantcomponent}, as, 
\begin{equation}
\mu_{A_3} = (1 - r_{A_3})(1 - g_{A_0}(u)),
\end{equation}
where,
\begin{equation} 
 u = 1 - r_{A_3} + r_{A_3} g_{A_1}(u),
\end{equation}
\begin{equation}
g_{A_0}(u) = \sum_{k=0}^{\infty} P_A(k) u^k 
\end{equation} and
\begin{equation}
g_{A_1}(z) = \displaystyle \sum_{k=0}^{\infty} q_{A_k}z^k
\end{equation}.

\subsection{Stage IV: Cascading failure in the power network}
\label{subsec:stageIV-powernetwork}
We now calculate the number of nodes in the power network which are connected to nodes not in the giant component in the communication network. 
The fraction of nodes which are removed because they have all their support links from the nodes not in the giant component of $N_A$, is given by, 
\begin{equation}
r_{B_4} = \displaystyle \sum_{\tilde{k}_B=0}^{\infty} \tilde{P}_B(\tilde{k}_B) (1 - \mu_{A_3})^{\tilde{k}_B}.
\end{equation}
 The giant component can be calculated as in Section \ref{sec-giantcomponent}. 

\subsection{Giant components and steady state conditions}
We will now calculate the size of the giant component at steady state. 
Let, $r_{A_{2n-1}}$ ($n \ge 1$) be the fraction of nodes in $N_A$ that are removed due to the removal 
of nodes in $N_B$ at stage $2n-2$.  
For $n=1$, the analysis is given in Section \ref{subsec:target-attack-communication-node}. 
Then, 
\begin{equation}
r_{A_{2n-1}} = \sum_{\tilde{k_A}=0}^{\infty}(1-\mu_{B_{2n-2}})\tilde{P}_A(\tilde{k}_A). 
\end{equation}

Proceeding similarly as above, the general expression for nodes for the fraction of nodes in the giant component at the $(2n-1)$-th stage in the communication network is 
given by, 
\begin{equation}
\mu_{A_{2n-1}} = (1 - r_{A_{2n-1}}) (1 - g_{A_0}(u)),
\end{equation}
where,
\begin{equation} 
 u = 1 - \phi_{A_{2n-1}} + \phi_{A_{2n-1}} g_{A_1}(u),
\end{equation}
\begin{equation}
g_{A_0}(u) = \sum_{k=0}^{\infty} P_A(k) u^k 
\end{equation} and
\begin{equation}
g_{A_1}(z) = \displaystyle \sum_{k=0}^{\infty} q_{A_k}z^k.
\end{equation}

Similarly, let, $r_{B_{2n}}$ be the fraction of nodes in $N_B$ that are removed due to the removal
of nodes in $N_A$ at stage $2n-1$. Then, 
\begin{equation}
r_{B_{2n}} = \displaystyle \sum_{\tilde{k}_B=0}^{\infty} \tilde{P}_B(\tilde{k}_B) (1 - \mu_{A_{2n-1}})^{\tilde{k}_B}
\end{equation}
The fraction of nodes in the giant component of $N_B$ at stage $2n$ is given by, 
\begin{equation}
\mu_{B_{2n}} = (1 -r_{B_{2n}}) (1 - g_{B_0}(u)),
\end{equation}
where,
\begin{equation} 
 u = 1 - \phi + \phi g_{A_1}(u),
\end{equation}
\begin{equation}
g_{B_0}(u) = \sum_{k=0}^{\infty} P_B(k) u^k 
\end{equation} and
\begin{equation}
g_{B_1}(z) = \displaystyle \sum_{k=0}^{\infty} q_{B_k}z^k
\end{equation}.

We arrive at a steady state when, 
\begin{eqnarray}
\mu_{A_{2n-1}} & = &\mu_{A_{2n+1}} = \mu_{A_{2n+3}} = \ldots\\
\mu_{B_{2n-2}} & = & \mu_{B_{2n}} = \mu_{B_{2n+2}} = \ldots
\end{eqnarray}

It is difficult to solve these systems of equations analytically. So, we generate the smart grid using different random graph models and simulate the effect of targeted and random attacks on these graphs. The results of this study is given in the next section. 

\subsection{Random Attacks}
\label{sec:random_attack}
In random attacks, the attacker chooses the nodes of a network, either uniformly at random, or according to a probability distribution defined on the nodes. If the network has $n$ nodes, the attacker chooses each node of the communication network with a probability of $\frac{1}{n}$ (for the uniformly at random case). This causes a cascading failure in the power network and the process is repeated.

\subsection{Adaptive Attacks}
\label{sec:adaptive_attack}
In adaptive attacks, the attacker deletes nodes iteratively in rounds, instead of all nodes at once. In each round, the attacker chooses a set of nodes to be compromised based on the new centrality measure and removes this set. It has been observed through experiments, that an adversary has an advantage while compromising nodes adaptively, compared to non-adaptive deletion. The paper \cite{wang2022adaptive} discusses some strategies for defending against adaptive attacks.

\section{Experimental results}
\label{sec:experimental}
\subsection{Experimental Set-up}
In order to simulate a smart grid, we use the network library \emph{igraph} \cite{GRAPH} on C. 
{\textcolor{black}{
Since previous studies \cite{PA13} have shown that power grids follow a power law degree distribution, we have considered the power network as SF network and compared it with ER networks.
} 
We consider two networks: the power network and the communication network, both of which are scale-free (SF) networks. 
For each communication node, an interlink is assigned by choosing a power node at random. 
We consider three types of attack on the communication network -- targeted, random and mixed (combination of the first two). In the random attack, we choose $x$ nodes uniformly at random from all the nodes without replacement. 
In the targeted attack, we choose $x$ nodes without replacement, such that the probability of choosing a node is proportional to the degree. For mixed attacks, we select half of the nodes for targeted attack and half of the nodes for random attack.
We evaluate the resilience of networks based on the size of the giant components and average path length under attack, in either of the networks (power and communication). 
Path length is an important parameter, because longer the path length larger are the communication overheads and delay in message transmission. 
We study targeted attacks, by compromising nodes based on degree, betweenness, and clustering coefficients. 
Finally, we study the effect of compromise by running the experiment 50 times for each input $x$. 
Every time the same graphs are considered. 

We compare our results with interdependent networks, with ER-ER combination instead of SF-SF combination. 
We also study adaptive attacks. 

\subsection{Experimental results, observations and inferences}
In Figure \ref{Smart Grid} to Figure \ref{Adaptive-between}, we present our experimental results. All experiments reported in Figure \ref{Comm-GiantComp} to Figure \ref{Power-path-all} are performed on communication networks with 2000 nodes and power networks with 1000 nodes. 

\begin{figure}[ht]
\centering
\includegraphics[scale=0.45]{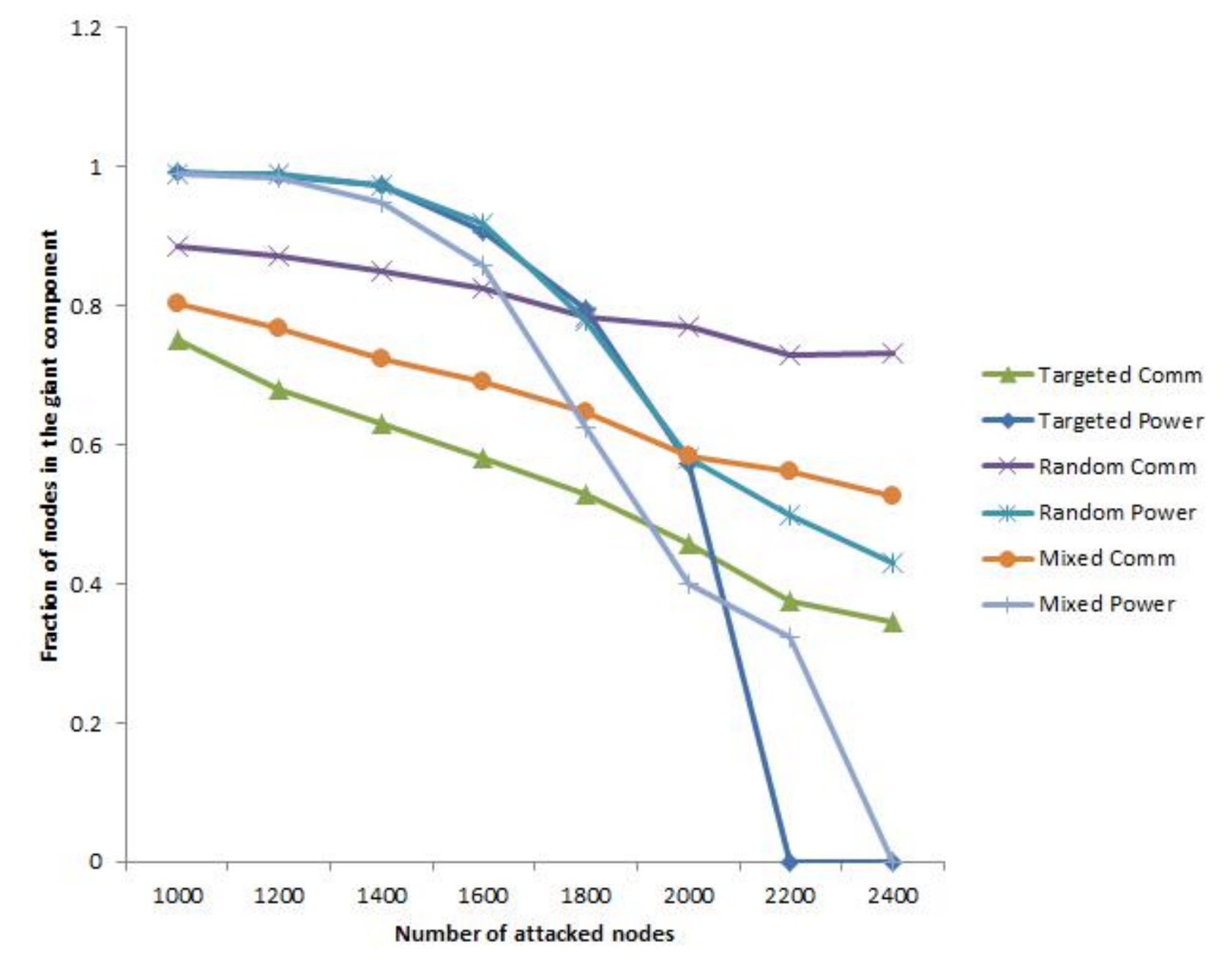}
\caption{Variation of giant component size with number of attacked nodes for targeted, random, and mixed attacks.}
\label{Smart Grid}
\end{figure}
 
\begin{figure}[ht]
\centering
\includegraphics[scale=0.45]{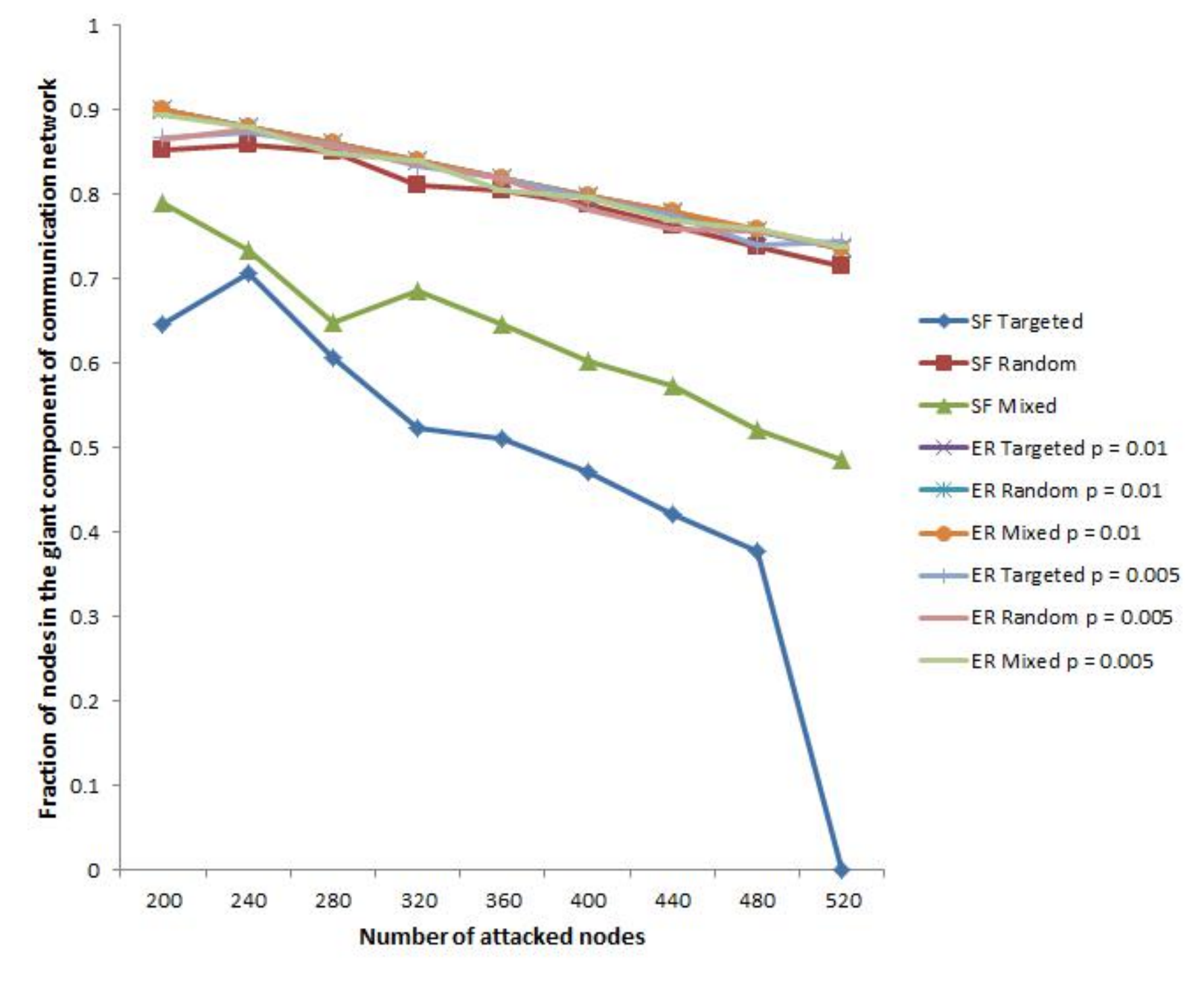}
\caption{Variation of giant component size with number of attacked nodes in the communication network for scale-free and Erdos-Renyi models.}
\label{Comm-GiantComp}
\end{figure}
 
\begin{figure}[ht]
\centering
\includegraphics[scale=0.45]{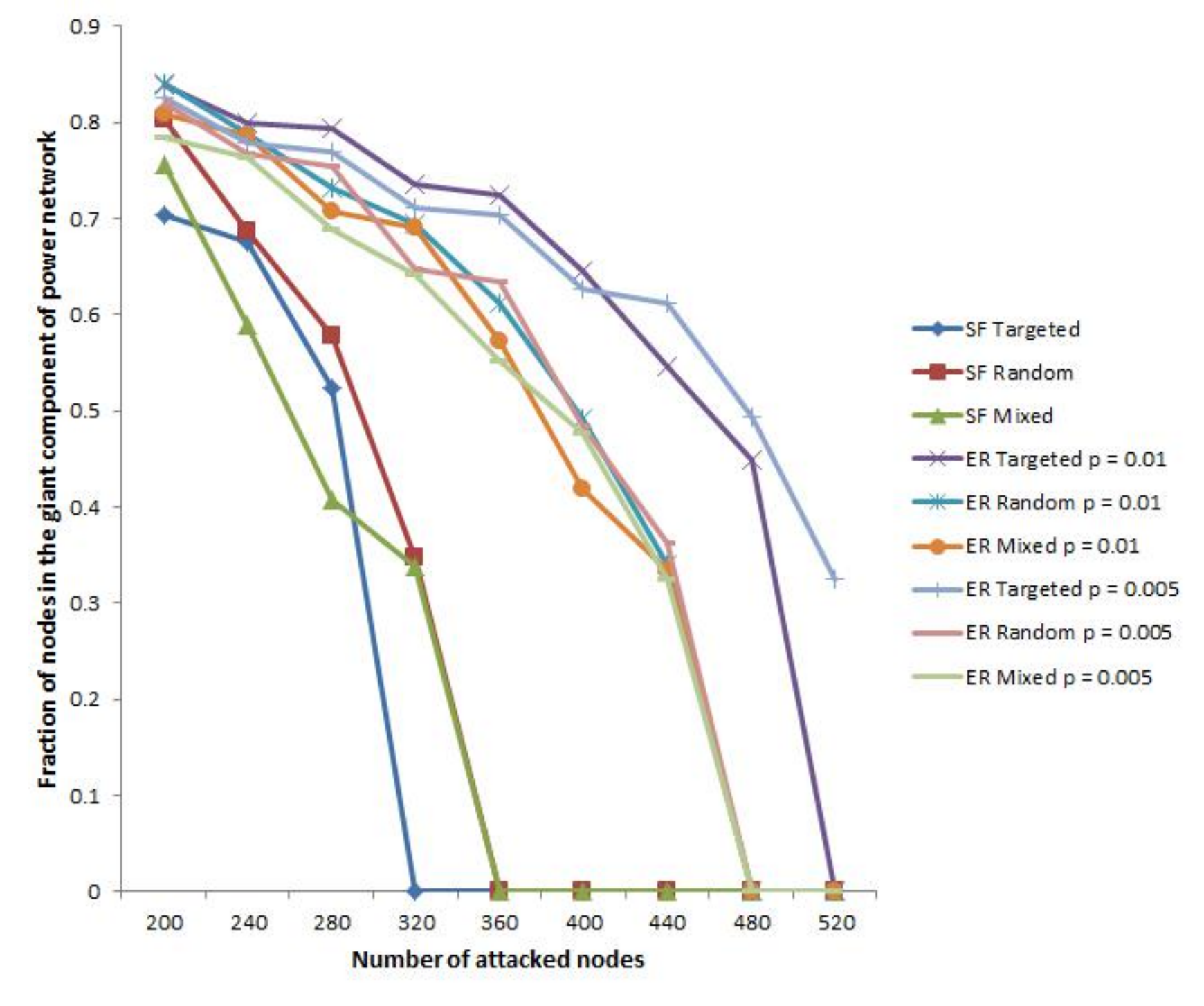}
\caption{Variation of giant component size with number of attacked nodes in the power network for scale-free and Erdos-Renyi models.}
\label{Power-GiantComp}
\end{figure}
 
In Figure \ref{Smart Grid}, the power network consists of 1,000 nodes and the communication network consists of 10,000 nodes. 
The communication/power network is generated as a scale-free network using a power-law degree distribution. 
We have plotted the size of the giant component (as a fraction of the size of the communication/power network) against the number of nodes attacked in the communication network. 
We observe that for a given value of the number of attacked nodes (only nodes in the communication network are attacked), 
the fraction of nodes in the giant component of the communication network is 
highest for random attacks and lowest for targeted attack (based on degree). 
The corresponding fraction for mixed attacks lies somewhere in the middle.
We also see that for the same fraction of nodes compromised, the giant component of the power network disintegrates faster for targeted attacks, 
compared to random attacks. We see that on compromising 2200 nodes, there is no giant component when targeted attack occurs, whereas giant component exists, under random attacks. 
This is expected, as attacking higher degree nodes result in a faster disintegration of the network, resulting in smaller components. 

In Figures \ref{Comm-GiantComp} and \ref{Power-GiantComp}, 
the communication network consists of 2000 nodes, whereas the power network consists of 1000 nodes.
We compare the results for a combination of SF-SF and ER-ER networks. 
In Figure \ref{Comm-GiantComp}, we have plotted the size of the giant component (as a fraction of the size of the communication network) 
against the number of attacked nodes. The targeted attack is based on the degree of nodes, i.e., high degree nodes are compromised with high probability. The communication network is generated using (i) a scale-free (SF) network 
using a power-law degree distribution, (ii) the Erdos-Renyi (ER) $G(n,p)$ model with $p = 0.01$, and (iii) the Erdos-Renyi 
(ER) $G(n,p)$ model with $p = 0.005$. 
Only nodes in the communication network are attacked. 
In Figure \ref{Power-GiantComp}, we have plotted the size of the giant component 
(as a fraction of the size of the power network) against the number of attacked nodes (based on degree). The power network is generated 
using the same models as mentioned above. From Figures \ref{Comm-GiantComp} and \ref{Power-GiantComp} we see that, 
for all the three types of attacks, the sizes of the giant components for ER graphs are comparable. 
The power and communication networks will be more fault tolerant to targeted attacks for Erdos-Renyi networks, compared to scale free networks.

\begin{figure}[ht]
\centering
\includegraphics[width = 3in]{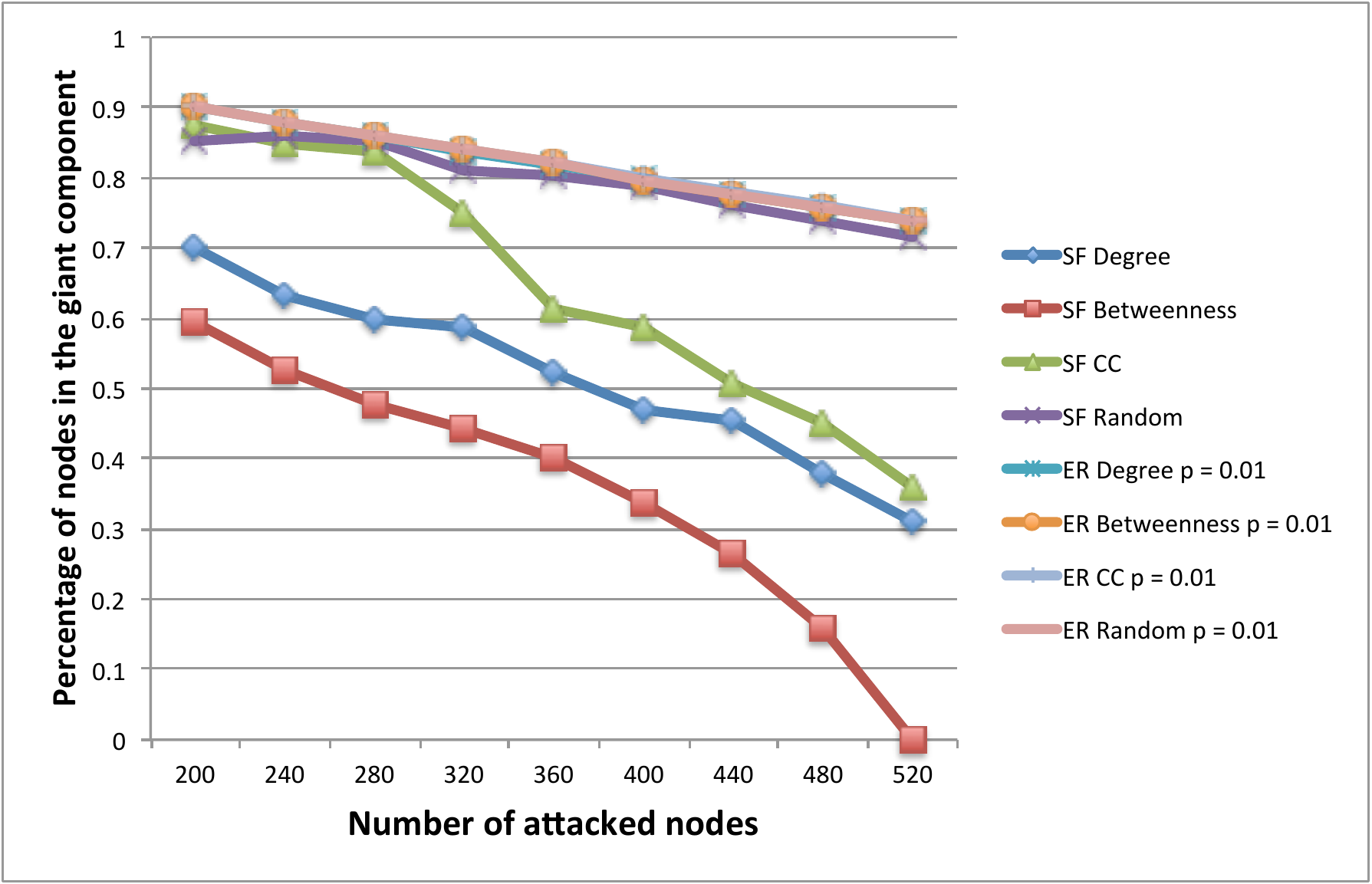}
\caption{Variation of giant component size of communication network with number of attacked nodes under different attack models, in the communication network for scale-free and Erdos-Renyi models.}
\label{Comm-GiantComp-all}
\end{figure}

\begin{figure}[ht]
\centering
\includegraphics[width = 3in]{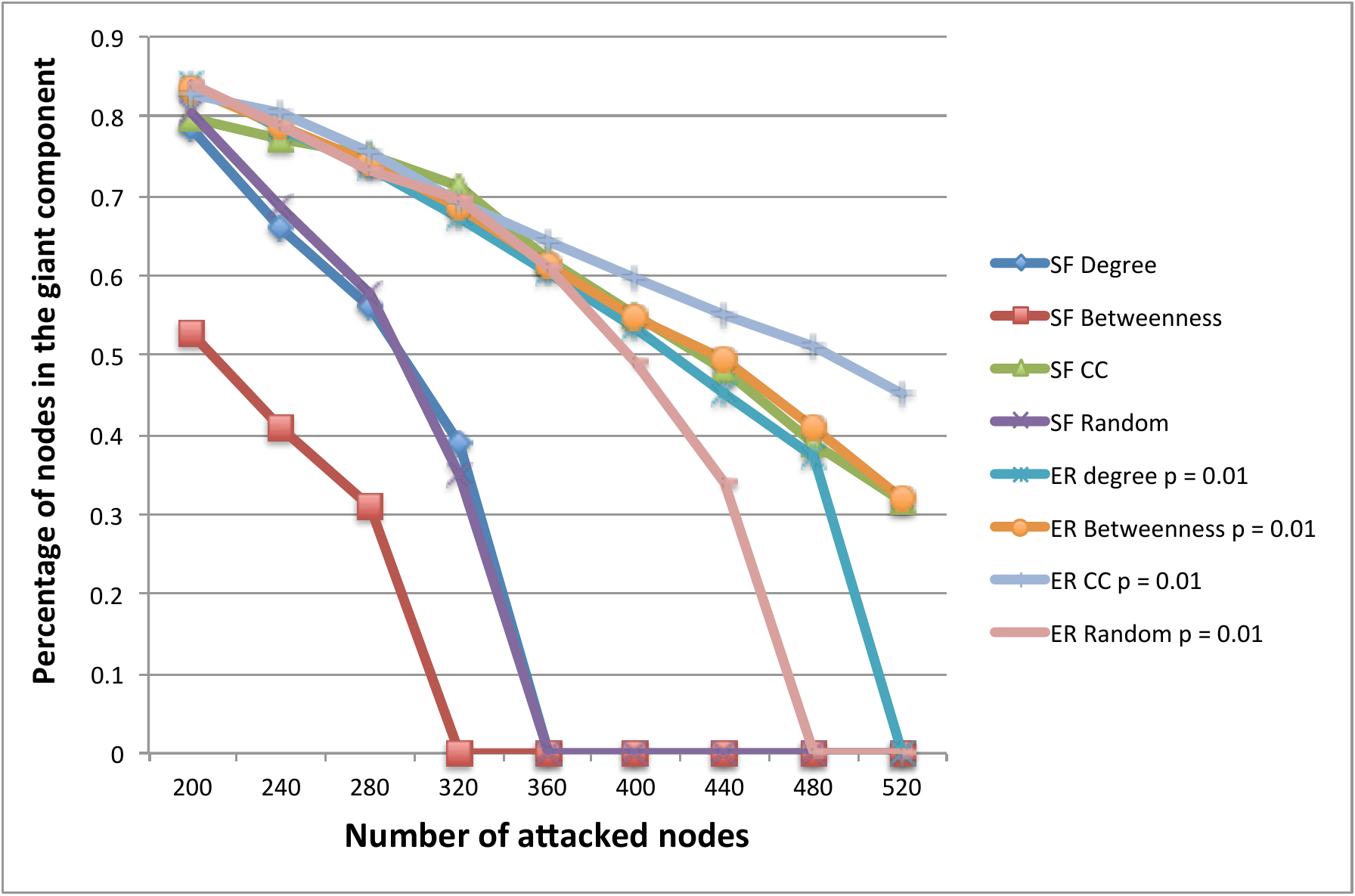}
\caption{Variation of giant component size of power network with number of attacked nodes in the power network for scale-free and Erdos-Renyi models.}
\label{Power-GiantComp-all}
\end{figure}

\begin{figure}[ht]
\centering
\includegraphics[width = 3in]{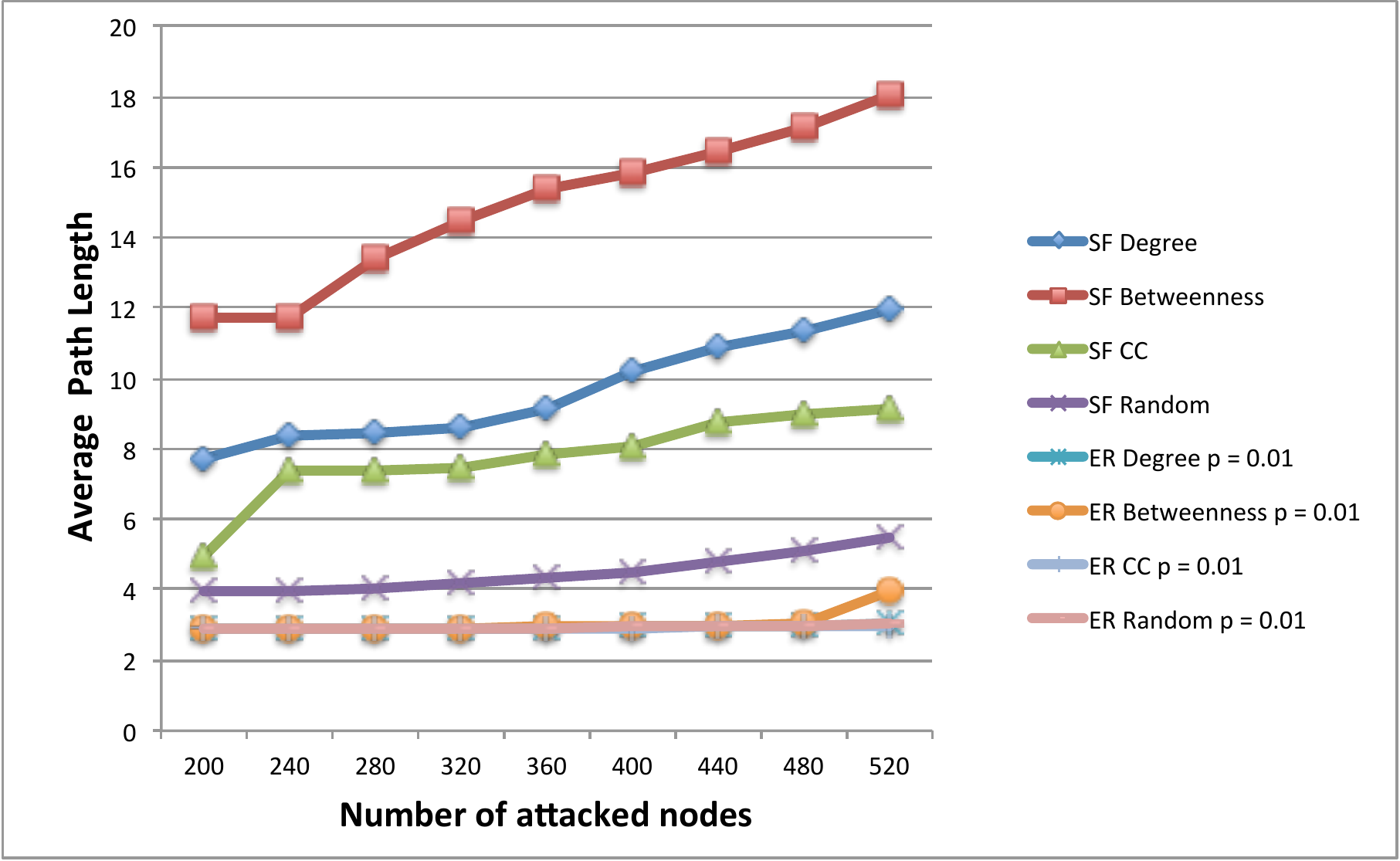}
\caption{Variation of average path length in communication network with number of attacked nodes in the network for scale-free and Erdos-Renyi models.}
\label{Comm-path-all}
\end{figure}

\begin{figure}[ht]
\centering
\includegraphics[width = 3in]{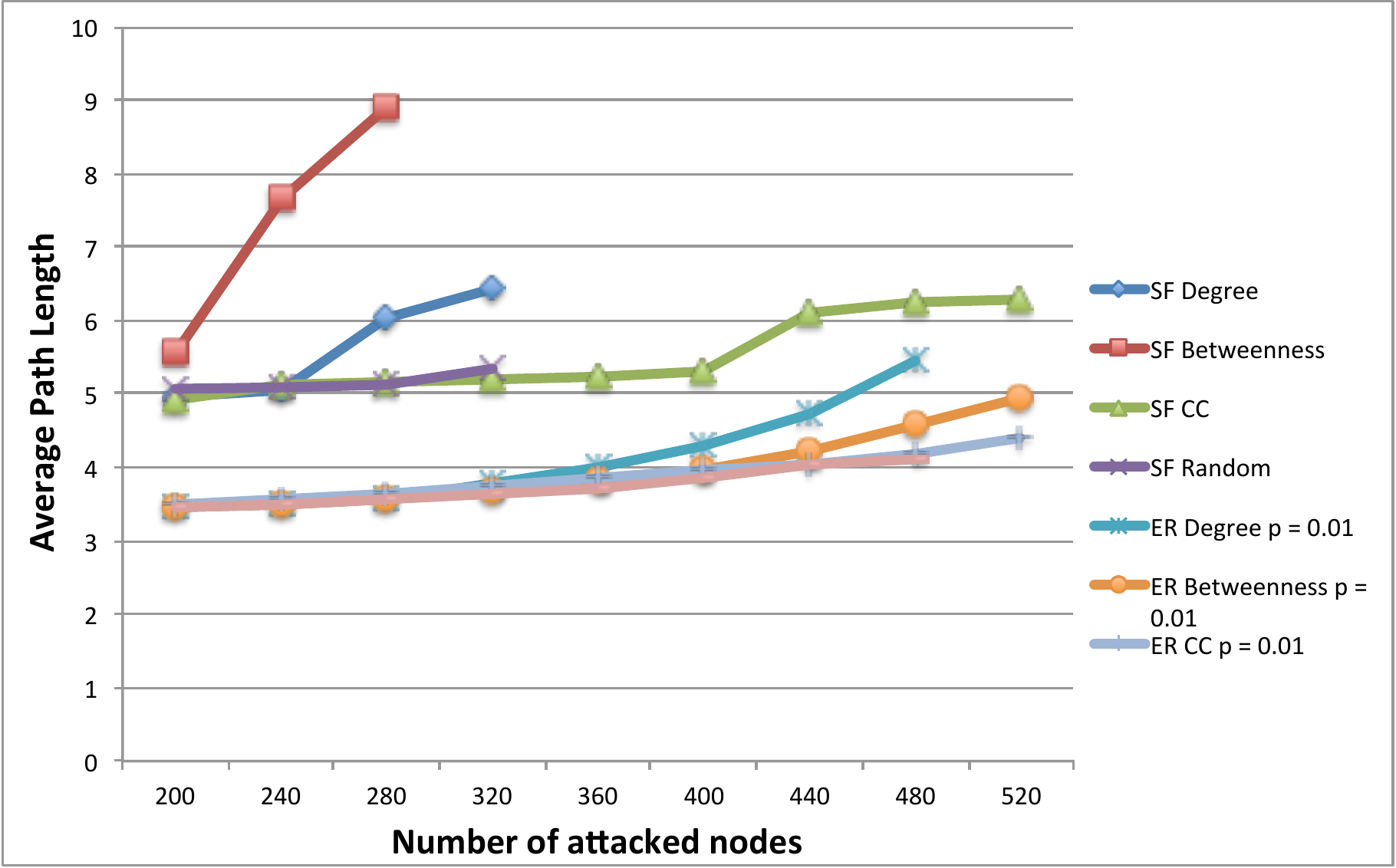}
\caption{Variation of average path length in power network with number of attacked nodes in the network for scale-free and Erdos-Renyi models.}
\label{Power-path-all}
\end{figure}

In Figures \ref{Comm-GiantComp-all} and \ref{Power-GiantComp-all}, 
the communication network again consists of 2000 nodes, whereas the power network consists of 1000 nodes. 
In Figure \ref{Comm-GiantComp-all}, we have plotted the size of the giant component (as a fraction of the size of the communication network) 
against the number of attacked nodes for various types of attacks. 
We have considered two different types of networks, Scale-free (SF) and Erdos-Renyi (ER) with $p = 0.01$. 
For the same SF networks, we have considered attack of nodes based on their (i) degree, (ii) betweenness, and
(iii) clustering coefficients, and compared them with random attacks. 
We have done a similar study for ER networks and considered the different types of attacks as above for the same network. 
We found that, to cause the maximum damage (get the smallest giant component) to a SF network, an adversary must delete high betweenness nodes with high probability. 
The next best strategy for an attacker is to compromise high degree nodes with high probability. The third best strategy is 
to compromise nodes with high clustering-coefficient with high probability. 
For ER networks, the attacker can use any attack strategy, because all of them give approximately the same results. In Figure \ref{Power-GiantComp-all}, we have plotted the size of the giant component (as a fraction of the size of the power network) 
against the number of attacked nodes for various types of attacks (based on degree, betweenness and clustering coefficients). Here we can see that for both scale-free and Erdos-Renyi networks, targeted attacks based on betweenness is the most effective strategy, while random attacks is the least effective from the point of view of an adversary.
{\textcolor{black}{
We observe that after targeted attacks, the size of the giant component in Erdos-Renyi networks can be twice as large as that in Scale-Free  networks. 
}

In Figures \ref{Comm-path-all} and \ref{Power-path-all}, 
the communication network again consists of 2000 nodes, whereas the power network consists of 1000 nodes. 
The average path length is a measure of how many hops a message must travel to reach the destination and should be minimized, in order
to reduce the communication overhead. 
In Figure \ref{Comm-path-all} we compare the average path length due to node compromise for the communication network.
We have considered only the average path length in the giant component, because that is the largest functional component in the network. 
For the same SF networks, we have considered following attack of nodes based on their (i) degree, (ii) betweenness, 
(iii) clustering coefficients, and compare them with random attacks. 
We have done a similar study for ER network and considered the different types of attacks as above for the same network. 
For SF network, we see that on compromising nodes
with probability proportional to the betweenness, the path average length increases the most. 
The average path length is the longest for case (ii), followed by (i) and (iii). 
For ER networks, the path length for all the three cases remain almost the same.
In Figure \ref{Power-path-all} we compare the average path length due to node compromise for the power network. In this case, the average path length is highest for SF networks when the attack is done based on betweenness and lowest for attack based on clustering coefficient. For ER graphs, the average path length is highest for attack based on degree and lowest for attack based on betweenness.
In Figure \ref{Power-path-all}, the average path length have not been plotted for node compromise beyond 320 nodes in SF-SF networks (for compromise based on degree/betweenness/clustering coefficients), because the giant component vanishes at this point. 

\begin{figure}[ht]
\centering
\includegraphics[scale=0.45]{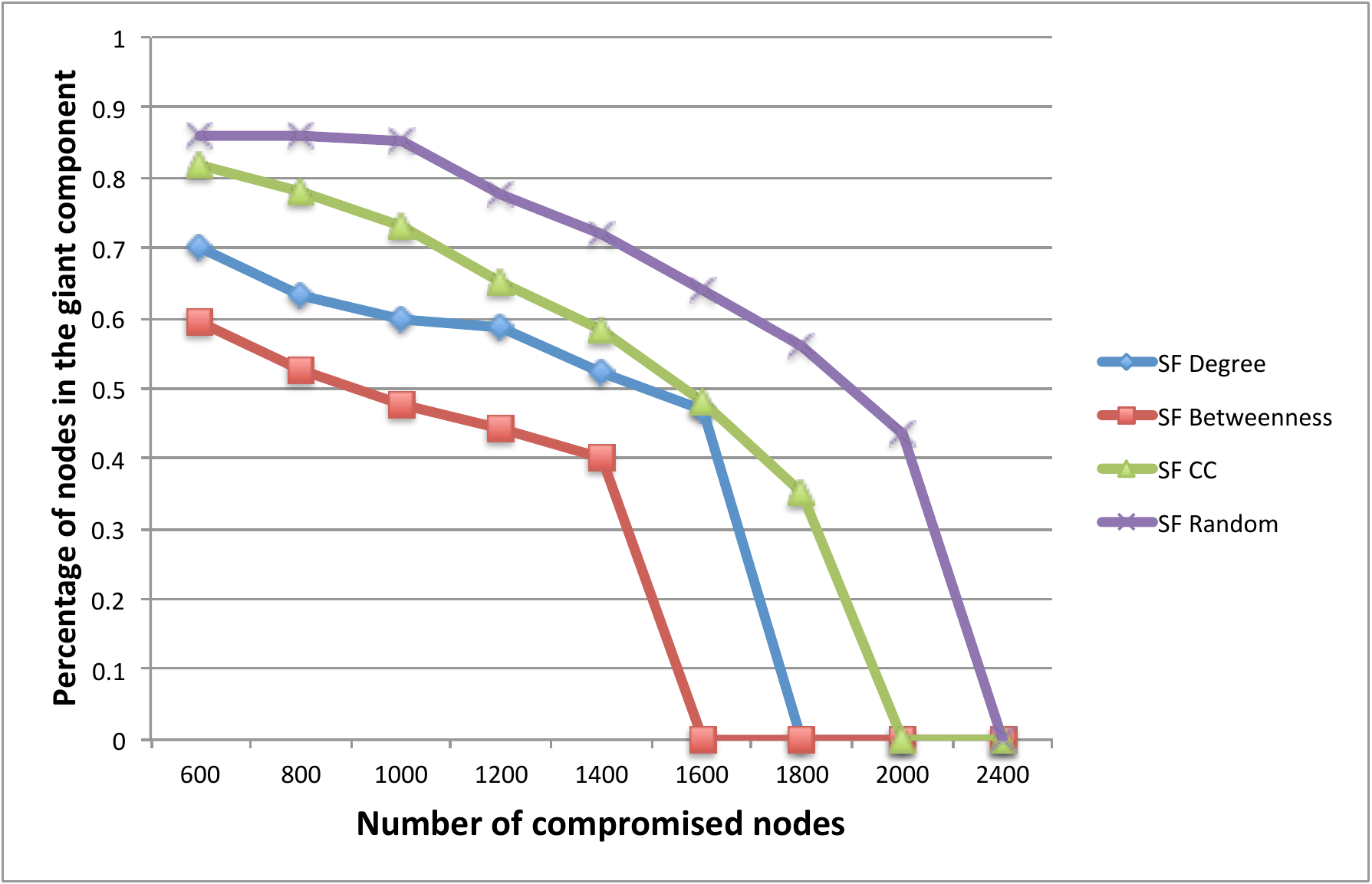}
\caption{Variation of giant component size with number of attacked nodes in the communication network for Western States power grid coupled with simulated communication network.}
\label{Real-Comm}
\end{figure}
 
\begin{figure}[ht]
\centering
\includegraphics[scale=0.45]{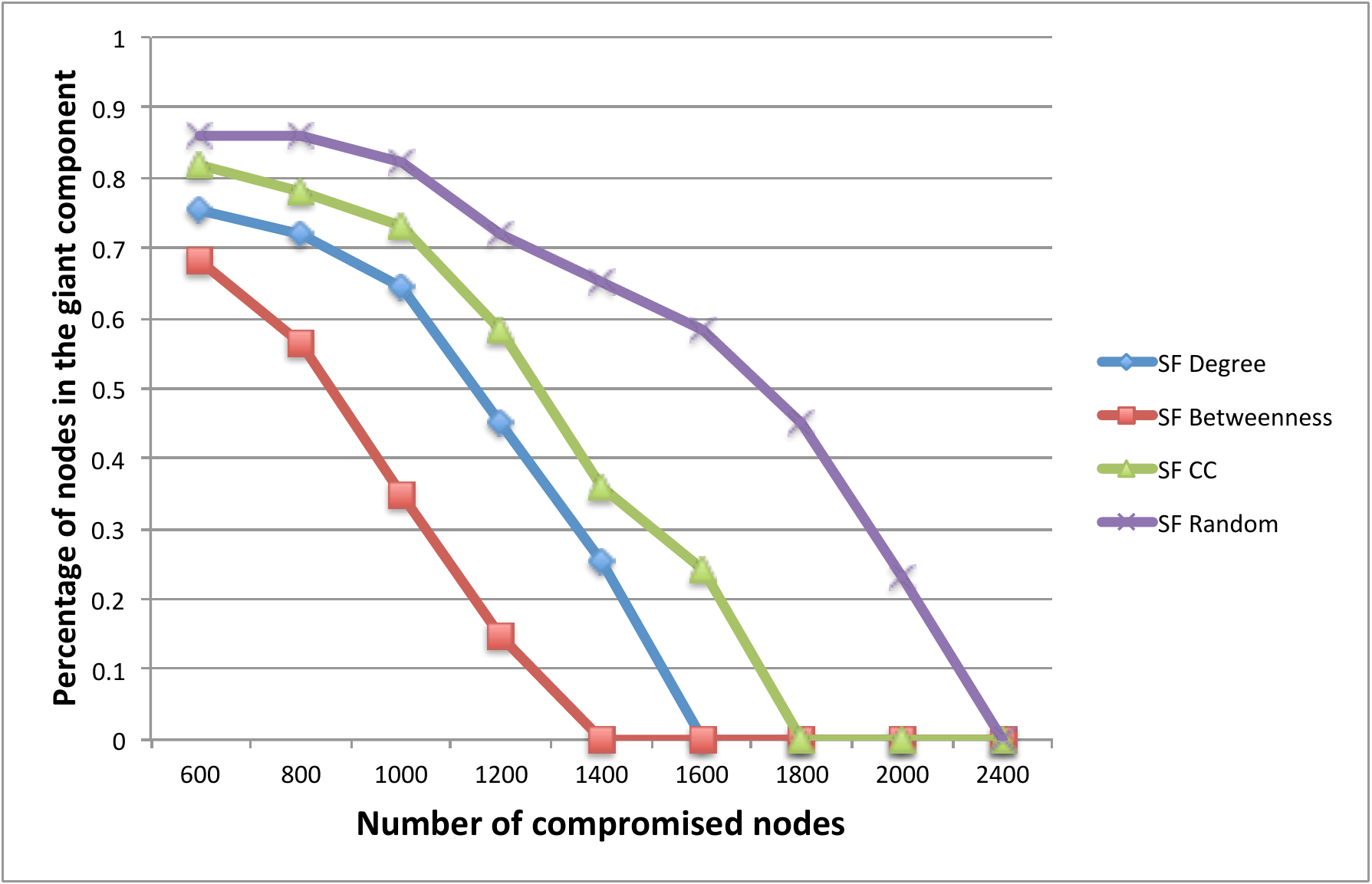}
\caption{Variation of giant component size with number of attacked nodes in the power network for Western States power grid coupled with simulated communication network.}
\label{Real-Power}
\end{figure}

{\textcolor{black}{
 We have carried out experiments with real power grid data for Western States power grid \cite{WS98}. We have coupled this network with synthetic SF communication network. The Western States power grid consists of 4941 nodes, whereas the communication network consists of 9000 nodes. We compare random attack with targeted attack based on degree, betweenness and clustering coefficient. As in the previous case, the best attack strategy for the attacker is to compromise nodes based on betweenness and the worst strategy is to compromise nodes randomly. The results are shown in Figures \ref{Real-Comm} and \ref{Real-Power}. 
}

\begin{figure}[ht]
\centering
\includegraphics[width = 3in]{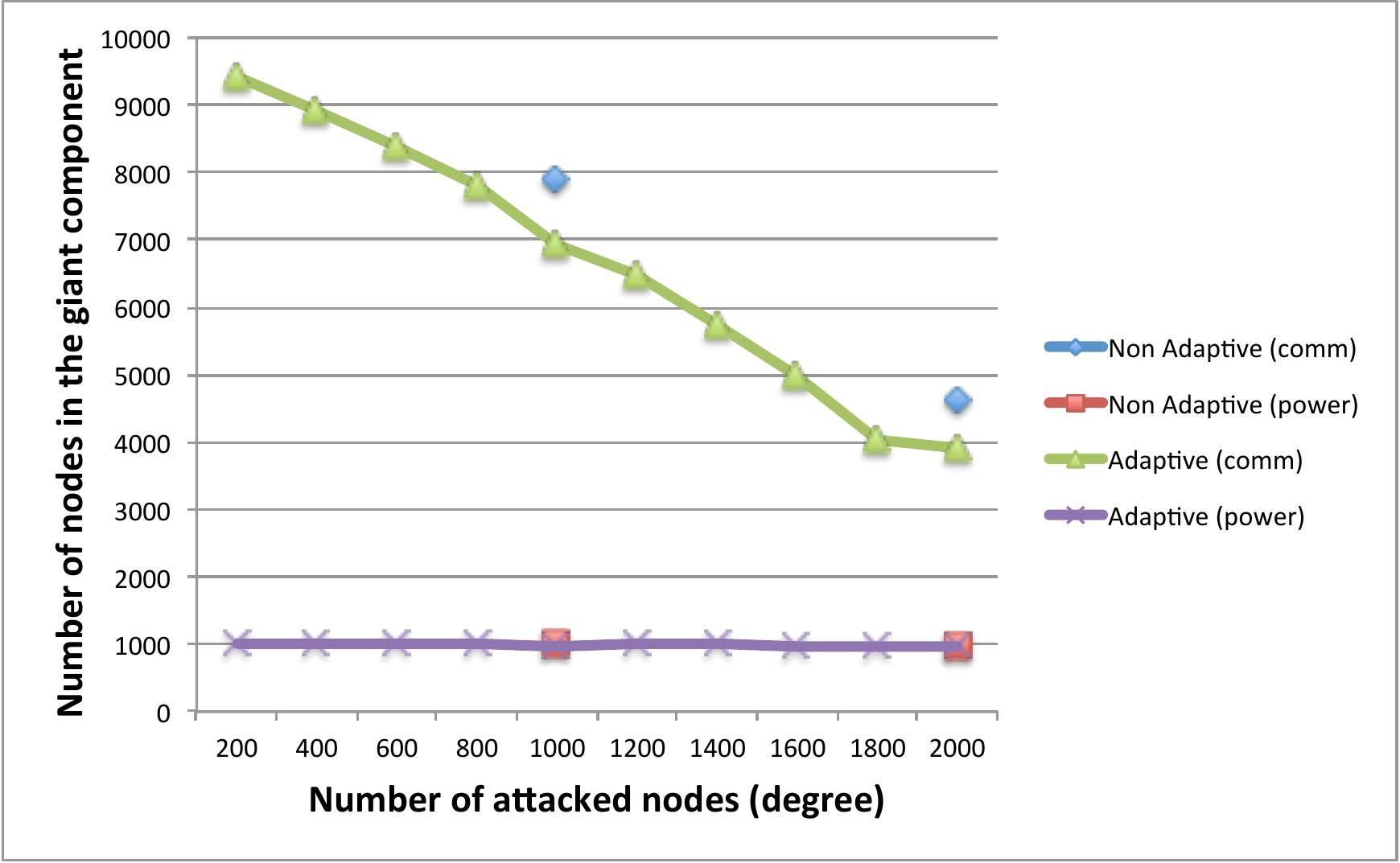}
\caption{Variation of giant component size of communication and power network with number of attacked nodes (based on degree) for adaptive and non-adaptive deletion.}
\label{Adaptive}
\end{figure}

\begin{figure}[ht]
\centering
\includegraphics[width = 3in]{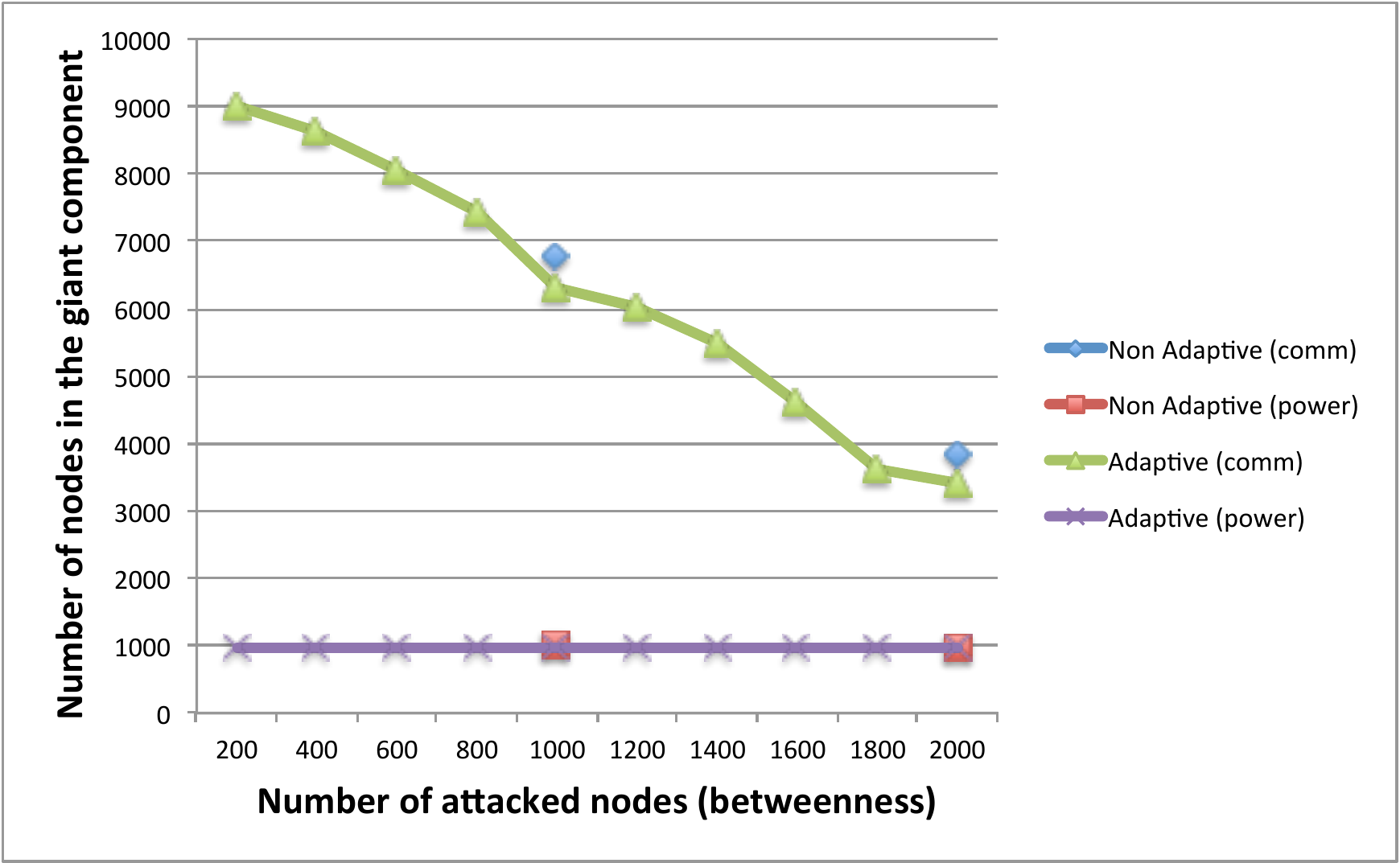}
\caption{Variation of giant component size of communication and power network with number of attacked nodes (based on betweenness) for adaptive and non-adaptive deletion.}
\label{Adaptive-between}
\end{figure}

In Figure \ref{Adaptive}, we present experimental results for adaptive attacks. 
We consider communication and power scale free networks for sizes 10000 and 1000 each. We compromise nodes in the following ways:
\begin{enumerate}
\item We consider adaptive attacks, in which we compromise 200 nodes at a time, every time calculating the new set of compromised nodes, with probability proportional to the degree at that instant.
\item We consider attacks where 1000 nodes are compromised together, with probability proportional to the degree. 
\end{enumerate}

In Figure \ref{Adaptive-between}, we compromise nodes selectively and adaptively based on the betweenness. Similar as above, the adaptive node compromise results in smaller giant component than non-adaptive node compromise. Also, compromising nodes selectively based on betweenness of nodes, results in smaller size of giant component compared to that based on degree (as we have seen before). For example, when 1000 nodes are compromised adaptively based on degree, the size of giant component in communication network is 6915, whereas compromising adaptively based on betweenness results in a giant component of 6310 nodes in the communication network.
When nodes are compromised adaptively based on clustering coefficient, the adversary has less advantage compared to compromising adaptively based on the degree.  

{\textcolor{black}{
\subsection{Strategies for defending smart grids}
Although, we have mostly talked about how an attacker can target vulnerable nodes in a network, we can use this to design a robust and fault-tolerant network. Understanding attack strategies is important to design good defense strategies. Since it is advantageous for an attacker to do targeted attacks on nodes with high betweenness centrality, guarding these nodes (power stations and communication centers) is vital for the reliability of a smart grid.
}}
{\textcolor{black}{
ER networks are more robust to attacks than SF networks.  
Even with the current security measures taken by Governments across the world,  cyber-security of critical infrastructure is not fool-proof. This paper stresses the importance of protecting the important nodes in the network, because of their vulnerabilities to attack. Though most of the power grid networks have a power law degree distribution (\cite{PA13}), it might be better to design the underlying network as an ER network. This will result in more robust networks, though at the cost of efficiency. 
}}

\section{Conclusion and future work}
\label{sec:conclusion}
In this paper, we have modeled the power and communication networks in a smart grid as two interdependent networks, and analyzed the cascading failure in smart grids for targeted attacks. 
{\textcolor{black}{This is one of the early works on targeted attacks and adaptive attacks on smart grids, which are modeled as interdependent networks.}}
We have given a mathematical expression for the size of the giant component when nodes are compromised. 
We have carried out extensive experiments to show that targeted attacks give an advantage to the adversary over random attacks and adaptive attacks are more effective than non-adaptive attacks.
 For targeted attacks, choosing a node with a probability proportional to its betweenness is more effective than doing so with degree or clustering coefficient.
We have shown than SF-SF networks are more prone to targeted attacks, than ER-ER networks. 
 
A challenging open problem is to obtain a closed-form solution for the size of the giant component from the mathematical analysis that we have presented.
Another important question is to present a good model of smart grids, which will be resilient to both random and targeted attacks. 
The structure of both the power and communication networks and the assignment of interlinks need to be studied. 
An important question is to find which network model and interconnection pattern will increase the resilience of the smart grid. 
Since the smart grid will be a component of the Internet of Things (IoT), a future direction of work is to propose a model, which will be resilient
to attacks and can disseminate information rapidly in the network.

\bibliographystyle{plain}
\bibliography{smartgrid}

\end{document}